\documentclass{article}

\usepackage{arxiv}

\usepackage[utf8]{inputenc} % allow utf-8 input
\usepackage[T1]{fontenc}    % use 8-bit T1 fonts
\usepackage{hyperref}       % hyperlinks
\usepackage{url}            % simple URL typesetting
\usepackage{booktabs}       % professional-quality tables
\usepackage{amsfonts}       % blackboard math symbols
\usepackage{nicefrac}       % compact symbols for 1/2, etc.
\usepackage{microtype}      % microtypography
\usepackage{lipsum}		% Can be removed after putting your text content
\usepackage{graphicx}
\usepackage{doi}

\usepackage{authblk}
\usepackage{siunitx}
\usepackage{amsmath}
\usepackage{amssymb}
\usepackage[normalem]{ulem}
\usepackage{dsfont}

\usepackage{tikz}

\usepackage{xargs}
\usepackage[colorinlistoftodos,prependcaption,textsize=tiny]{todonotes}
\newcommandx{\cJP}[2][1=]{}%{\todo[linecolor=blue,backgroundcolor=blue!25,bordercolor=blue,#1]{#2}}

\author[1,2]{Awal Awal}
\author[2]{Jan Hetzel}
\author[2,3]{Ralf Gebel}
\author[1,3]{Jörg Pretz}
\affil[1]{RWTH Aachen University, 52056 Aachen, Germany}
\affil[2]{GSI Helmholtzzentrum für Schwerionenforschung GmbH, 64291 Darmstadt, Germany}
\affil[3]{Forschungszentrum J\"ulich GmbH, 52425 J\"ulich, Germany}

\date{June 18, 2024}

\renewcommand{\headeright}{Injection Optimization at Particle Accelerators via RL}
\renewcommand{\undertitle}{}
\renewcommand{\shorttitle}{}

\hypersetup{
pdftitle={Injection Optimization at Particle Accelerators via Reinforcement Learning: From Simulation to Real-World Application},
pdfsubject={},
pdfauthor={A. Awal, J. Hetzel, R. Gebel, J. Pretz},
pdfkeywords={machine learning, reinforcement learning, particle accelerator, Cooler Synchrotron COSY},
}

\renewcommand{\sout}[1]{}

\title{Injection Optimization at Particle Accelerators via Reinforcement Learning: From Simulation to Real-World Application}

\begin{document}
\maketitle

\begin{abstract}
Optimizing the injection process in particle accelerators is crucial for enhancing beam quality and operational efficiency. This paper presents a framework for utilizing Reinforcement Learning (RL) to optimize the injection process at accelerator facilities. By framing the optimization challenge as an RL problem, we developed an agent capable of dynamically aligning the beam's transverse space with desired targets. Our methodology leverages the Soft Actor-Critic algorithm, enhanced with domain randomization and dense neural networks, to train the agent in simulated environments with varying dynamics promoting it to learn a generalized robust policy. The agent was evaluated in live runs at the Cooler Synchrotron COSY and it has successfully optimized the beam cross-section reaching human operator level but in notably less time. An empirical study further validated the importance of each architecture component in achieving a robust and generalized optimization strategy. The results demonstrate the potential of RL in automating and improving optimization tasks at particle acceleration facilities. 
\end{abstract}

% keywords can be removed

\section{Introduction}
\label{sec:intro}
The field of accelerator physics has seen a growing interest in leveraging machine learning techniques to enhance the operation and efficiency of particle accelerators. Among the machine learning techniques, Reinforcement Learning (RL) has emerged as a powerful tool for optimizing complex systems~\cite{tokamak}. 

Particle accelerators are complex machines designed to accelerate charged particles to desired energies for a variety of applications ranging from basic research in physics to applied sciences and medical treatments~\cite{acc-med, acc-med2}. The Cooler Synchrotron COSY~\cite{cosy1,cosy2}, located at Forschungszentrum Jülich in Germany, is an accelerator facility primarily focused on hadron physics research\cite{cosy-axion,cosy-exotic-particle,cosy_legacy}. Optimizing the injection process, which involves transferring particles into the accelerator's storage ring, is a non-trivial challenge. This process is critical for ensuring high-quality beam properties, such as intensity and stability, which directly impact the effectiveness and precision of experiments conducted at accelerator facilities~\cite{acc-challenges1, acc-challenges2}.

Integrating machine learning methods into accelerator facilities has been a subject of increasing interest within the scientific community. Several studies have demonstrated the potential of machine learning techniques in improving various aspects of accelerator operations. For instance, machine learning methods have been employed for beam diagnostics~\cite{diagnostic-ebeam, diagnostic-xray, diagnostic-longitudinal}, optimization and control tasks~\cite{bo-free-electron,Scheinker_2018, Leemann_2019, Scheinker_2021}, and anomaly detection~\cite{beam-anomaly1, beam-anomaly2}. These works, among others, highlight the versatility and effectiveness of machine learning approaches in addressing the challenges faced by accelerator facilities. 
Within the domain of utilizing RL in accelerator operations, several studies utilized RL methods to optimize the beam in simulation, showing their promising potentials~\cite{pang-2020, new-rl-sim, new-rl-sim2}. Recent studies have managed to run RL agents successfully in live runs, including optimizing a beam with linear settings~\cite{PhysRevAccelBeams.23.124801}, employing an RL agent that is trained only in simulation to optimize the beam at a real machine~\cite{new-rl-real, new-rl-real2}, and in addition optimize a beam with non-linear settings~\cite{new-rl-real2}. These recent methods, however, are either limited to linear optimization problems or low dimensional optimization in comparison to the IBL at COSY. Furthermore, achieving operator-level optimization remains a challenging task requiring complex RL methods that cannot be trained in live settings due to their need for a large amount of training. Achieving this level of optimization from simulation training only requires a more comprehensive approach in RL.

The optimization of the injection process in particle accelerators, such as the one at COSY, presents a machine learning challenge that is impractical for traditional supervised machine learning methods. Supervised learning depends on a labeled dataset where the correct output for each input is known and used to train the model. This is not the case in the context of injection optimization as the optimal actions are not readily available or easily computed due to the highly dynamic and complex nature of particle accelerators. In addition, there are no expert samples of optimal actions and it is difficult to build one within the domain of injection optimization because the optimal settings can vary significantly based on the specific experimental setup and objectives. If sufficient expert samples of optimal actions are available then other machine learning methods like imitation learning~\cite{imitation-learning} and inverse RL~\cite{inverse-rl} might be considered. 

RL is well-suited for this type of optimization tasks as it does not require labeled examples or expert samples. RL agents learn optimal policies through trial and error by interacting with the environment and are guided by a reward function that reflects the quality of the actions based on their outcomes. This approach allows RL agents to autonomously discover and refine strategies for optimizing the injection process~\cite{rl-intro}.

\begin{figure}[!t]
   \centering
   \includegraphics*[width=.8\columnwidth]{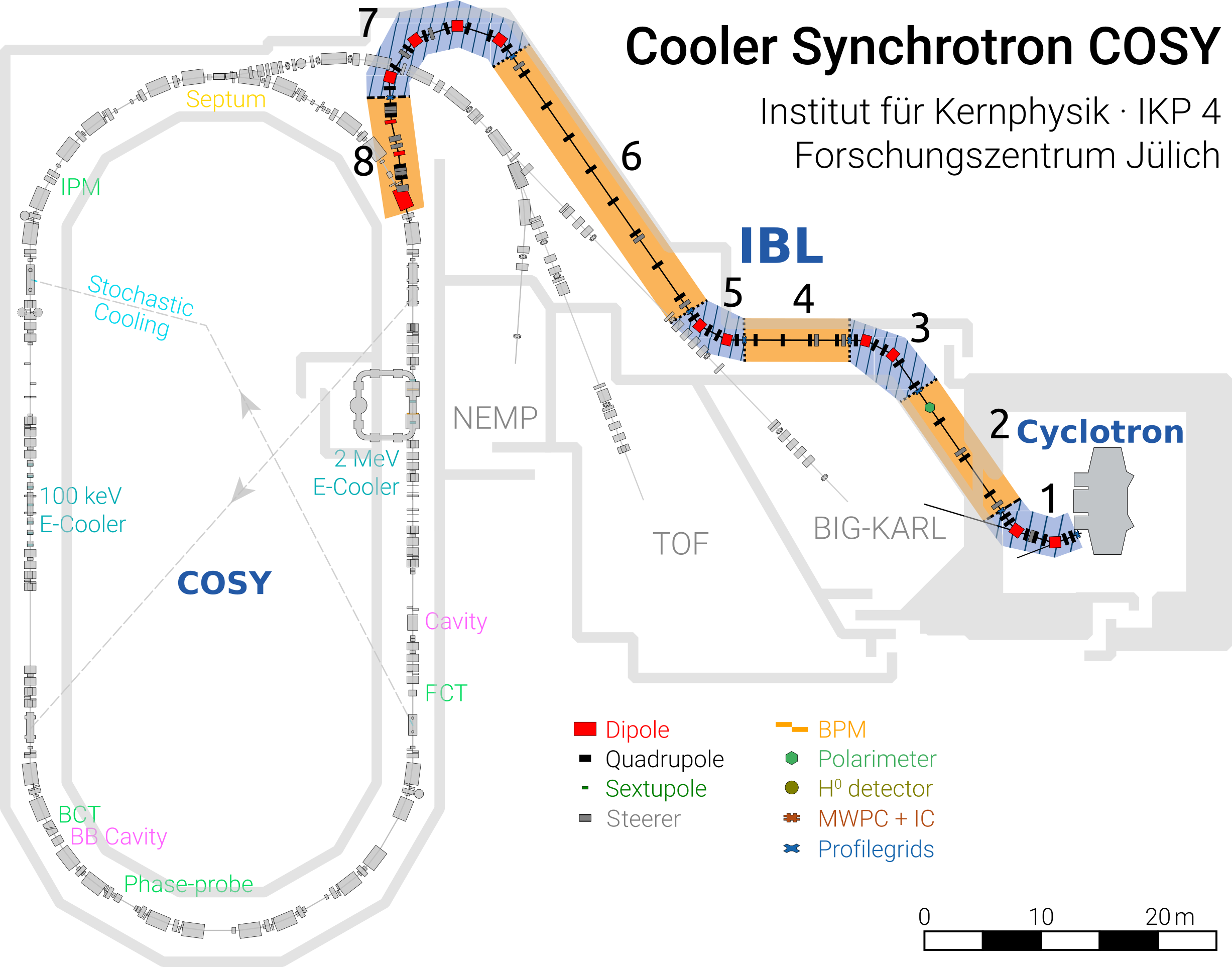}
   \caption{The COSY facility at Forschungszentrum Jülich (FZJ). Illustrated are the cyclotron (right), the cooler synchrotron COSY (left), and the interconnecting injection beam line IBL. For the latter, its division into sections is indicated with colors and numerals.}
   \label{fig:IBL_and_COSY}
\end{figure}

Numerical optimization methods such as Bayesian optimization and evolution strategies are popular methods within the domain of optimizations in accelerator facilities due to their flexibility and adaptability~\cite{bo, bo-free-electron, Appel:208544, Catani:453324}. These methods are easier to implement and are well-suited for scenarios where a simulation model is not available or when quick adjustments are needed for different optimization targets. However, they tend to require more time to converge with less consistency, especially in high-dimensional cases. On the other hand, RL, while requiring more initial investment in terms of computational resources and time, provides efficient, fast, and consistent optimization performance through targeted exploration by learning the system dynamics.

Building on our previous work and interest~\cite{bo}, this research introduces a framework for the application of RL in accelerator facilities and applies it to optimize the injection process at COSY. By framing the optimization challenge as an RL problem, we train an agent to make data-driven decisions that improve the injection efficiency. The RL agent is capable of adapting to varying conditions, learning from interactions with the environment to optimize the beam's properties dynamically. The actions of RL agents are aligned and tailored to the specific domain they are trained on. This is because they build an internal concept of their environment and the potential consequences of their actions~\cite{othello}. This paper is based on the thesis~\cite{awal-thesis} which contains a more detailed description.

\section{Beam and Injection Optimization at The Cooler Synchrotron COSY Facility}
\label{sec:cosy}

COSY accelerates protons or deuterons and is equipped to handle polarized and unpolarized ions. The facility, including the cyclotron and the injection beam line (IBL), is used in a wide range of experimental research. The IBL at COSY is specifically designed for the transfer of negatively charged hydrogen and deuteron ions from the cyclotron JULIC~\cite{julic} to the synchrotron. At the point where the ions are injected into COSY, the \emph{injection point}, the electrons are stripped off via a stripping foil. During injection, the acceptance of COSY is filled with particles via multi-turn phase space painting. Subsequently, the optimal region in phase space for the incoming beam during injection will be called \emph{injection acceptance}. The IBL's functionality is crucial, as it directly influences the intensity and quality of the beam delivered to COSY.

\subsection{The Injection Beam Line}
\label{subsec:ibl}
The IBL at COSY, spanning a length of \SI{94.15}{\meter}, is a transfer beam line designed for the efficient transfer of ions, see Figure~\ref{fig:IBL_and_COSY}. It is logically partitioned into 8 sections consisting of 15 quadrupole magnet families and 28 steerers, among other components, to guide and shape the beam into COSY. The IBL’s design allows for precise control over the beam, which is essential for achieving optimal performance in beam injection and experimentation. Figure~\ref{fig:IBL_and_COSY} illustrates the layout of the COSY facility, highlighting the IBL and its various sections.

\begin{figure}[!htb]
   \centering
   \includegraphics*[width=.95\columnwidth]{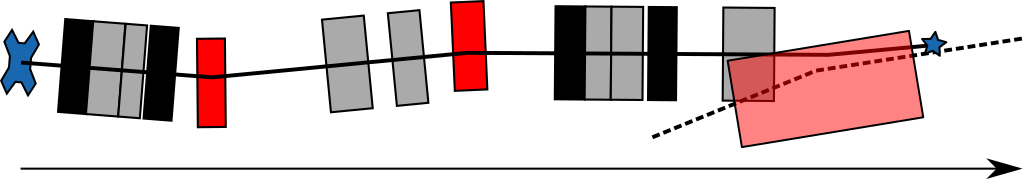}
   \caption{A sketch of section~8 of the IBL showing the central path and the different components along the beam line. Dipoles are colored in red, quadrupoles are colored in black, and steerers are colored in gray. The injection point is marked with a blue star, where the charge exchange foil is placed. The trajectory of the $p^-$ beam through the IBL is indicated by the solid black line. The dashed line indicates the central orbit through the synchrotron COSY.}
   \label{fig:sec8}
\end{figure}

Section 8, the last section of the IBL, is composed of 4 quadrupoles and 7 steerers and has high importance, see Figure~\ref{fig:sec8}. The task of this section is to match the transferred beam from the preceding sections with the injection acceptance of COSY. The focus of the RL approach is to optimize this section due to its vital and direct role in controlling the final phase of the beam injection process and the subsequent impact on the beam intensity and quality inside COSY.

\subsection{Injection Optimization}
\label{subsec:optimization}
The optimization of the IBL is a complex non-linear problem, primarily due to the multi-dimensional nature of beam dynamics and the complex design of the IBL. The objective is to maximize the beam intensity inside COSY while ensuring minimal setup time. The optimization process involves adjusting the parameters of the IBL components, particularly the strengths of the quadrupole and steerer magnets, to fine-tune the beam's trajectory and properties. The ultimate goal of this optimization is to align the phase space of the injected beam with the injection acceptance of the storage ring to ensure a stable and high intensity beam. A key challenge in the optimisation is, that the beam parameters at the beginning of  the IBL vary. This is in particular related to the cyclotron and its magnetic field, which drifts over time and thus influences the beam. This can even lead to beams where the cross section differs from a shape wich can be estimated by a Gaussian and offers multiple clusters of particles instead. Given the complexity of this system which only can be set up by experts, a common mode of operation at COSY is to treat the beam from the cyclotron as a black box and make adjustments to the IBL to compensate for the unknown beam parameters. Currently, the optimization process involves adjusting the magnets manually to maximize the beam current inside COSY, a process that is time-consuming and lacks consistency. Automating this process is of high significance to ensure minimal setup time and higher availability for the experiments. An alternative approach is to measure the parameters of the beam at the beginning of the IBL, as described in \cite{inj-opti-cosy-benat}. As the measurement of the parameters to the necessary precision is lengthy as well, this approach is usually not followed in standard operation.
%summary of procedure and idea

The automation of the injection optimization, as described in the following, is based on the assumption that the optimal beam occupation of the phase space at the charge exchange foil for injection is known and can be restored by an appropriate setting of the IBL. In the topical approach the task of finding this setting is divided in two sub tasks: The first is to transport the beam from the beginning of the IBL to the beginning of section eight without losing intensity. A possible approach to achieve this is Bayesian optimisation as it is described in \cite{bo}. The second task is then to use the elements in section eight to match the desired parameters at the injection point of COSY. Our approach is to use an RL agent to carry out the latter task. As in this approach the beam parameters at the beginning of the IBL are unknown and the first step only ensures full transmission of the beam, the beam parameters of the beginning of section eight are to be regarded as unknown as well. Dividing the optimisation in two tasks is chosen to reduce the number of free parameters for the RL agent to a reasonable amount while keeping the necessary degrees of freedom for a successful optimization. At COSY we have the opportunity to record the position and the cross section of the beam at the end of the IBL with a fluorescent screen. As this gives access to the spacial components only, the goal of the RL agent is to match the transverse position and spread of the beam at this location to an operator given specification. For testing and demonstration of the capabilities of the RL agent the specification may differ from the optimal settings for injection. The RL agent is trained in simulation only prior to its application to the real machine. The experimental demonstration includes the described matching to the user's specifications as well as an investigation of the influence of several architecture components on the result.

\section{Theoretical Background}
\label{sec:theory}
In this section, we introduce the theoretical foundations of reinforcement learning, along with the concepts of Partially Observable Markov Decision Process (POMDP) and domain randomization. These concepts are essential for understanding the RL framework applied in this research. A comprehensive introduction to RL can be found in \cite{rl-intro-2} and a detailed description is given in \cite{rl-intro}.

\subsection{Reinforcement Learning}
\label{subsec:rl}
Reinforcement Learning (RL) is the core principle of our approach to optimize the IBL at COSY. The general setup of RL involves an agent interacting with its environment at a sequence of time steps to maximize cumulative rewards. In this setup, the environment's state at each time step $t$ is represented by $s_t \in \mathcal{S}$. Assuming full observability of the state, the agent's policy $\pi(a | s)$ dictates the probability distribution over actions $a \in \mathcal{A}$ given a state $s$. The agent then receives a reward signal $r_t = r(s_t, a_t)$ from the environment, providing feedback on the desirability of the action. The agent's objective is to maximize the expected return over a horizon, which is the sum of discounted rewards obtained during an episode, expressed as:

\begin{equation}
    R_t = \sum_{k = t}^{T} \gamma^{k - t} r_{k} \, ,
\end{equation}

where $\gamma \in [0, 1]$ is the discount factor which balances the immediate and future rewards, and $T$ is the horizon of each episode. 

The state-value function $V^\pi(s)$ is the expected return over the horizon, starting from state $s$ and following a policy $\pi$. The state-value function $V^\pi(s)$ is defined as:

\begin{equation}
    V^\pi(s) = \mathbb{E}_{\tau \sim p(\tau | \pi)}\left[ \sum_{k = t}^{T} \gamma^{k - t} r_{k} | S_t = s \right] \, ,
\end{equation}

where $\tau = (s_0, a_0, s_1, ..., a_{T - 1}, s_T)$ denotes a trajectory through the state-action space and $p(\tau | \pi)$ is the trajectory's probability under policy $\pi$. The action-value function $Q^\pi(s, a)$, also known as the Q-function, representing the expected return when starting in state $s$ and taking action $a$ then following policy $\pi$, is closely related to $V^\pi(s_t)$ and is defined as:
\begin{equation}
    Q^\pi(s, a) = \mathbb{E}_{\tau \sim p(\tau | \pi)}\left[ \sum_{k = t}^{T} \gamma^{k - t} r_{k} | S_t = s, A_t = a \right].
\end{equation}

The state-value function $V^\pi(s_t)$ can be expressed in terms of the action-value function $Q^\pi(s, a)$ as:

\begin{equation}
    V^\pi(s) = \mathbb{E}_{a \sim \pi}\left[ Q^\pi(s, a) | S_t = s \right].
\end{equation}

$J(\pi)$ represents the expected return by following a certain policy $\pi$. This measure indicates how good a policy is and maximizing it results in improving the policy to achieve the optimal policy $\pi^*$. The expected return of a policy $J(\pi)$ is defined as:

\begin{equation}
    J(\pi) = \mathbb{E}[R_0 | \pi] = \mathbb{E}_{\tau \sim p(\tau | \pi)} \left[ \sum_{k = 0}^{T} \gamma^k r(s_k, a_k) \right]. %\quad , \, r_k \equiv r(s_k,a_k) \, ,
\end{equation}

The next state is dictated by the state transition function $P(s_{t + 1} | s_t, a_t)$ which is determined by the dynamics of the environment~\cite{rl-intro}. Contemporary methods in RL utilize a key concept in RL which is the Bellman equation~\cite{Bellman:DynamicProgramming}. It provides a recursive decomposition for the value function. 
The Bellman equation for $Q^\pi(s)$ is given as:
\begin{equation}
    Q^\pi(s, a) = \mathbb{E}\left[ r(s,a) + \gamma V^\pi(s_{t+1}) | S_t = s, A_t = a \right].
\end{equation}

The objective in RL is to find the optimal policy $\pi^*$ that maximizes the expected return from any initial state. This is often framed as maximizing the state-value function $V^\pi$ or the action-value function $Q^\pi$ for all states $s \in \mathcal{S}$. Mathematically, the objective is to find $\pi^*$ such that:
\begin{equation}
    \pi^* = \arg\max_\pi \mathbb{E}_{s \sim S}\left[V^\pi(s)\right].
\end{equation}

Table~\ref{tab:definitions} shows an overview of the variables used in the equations and their definitions.

\begin{table}[ht]
    \centering
    \begin{tabular}{ll}
    \hline
    \hline
    Variable & Meaning \\
    \hline
    $t$ & Time step \\
    $\tau$ & Trajectory \\
    $a$ & Action \\
    $s$ & State \\
    $o$ & Observation \\
    $h$ & History \\
    $\pi$ & Policy \\
    $\gamma$ & Discount factor \\
    $r$ & Reward \\
    $R_t$ & Future expected rewards from time step $t$ \\
    $J(\pi)$ & Expected return of a policy $\pi$\\
    $Q(s,a)$ & Action-value function \\
    $V(s)$ & State-values function \\
    $\rho$ & Environment dynamics vector\\
    $g$ & The goal of the optimization \\
    \hline
    \hline
    \end{tabular}
    \caption{Definitions of variables used throughout the paper.}
    \label{tab:definitions}
\end{table}

\subsection{Partially Observable Markov Decision Processes}
\label{subsec:pomdp}
RL methods are primarily designed to solve decision-making problems formulated as Markov Decision Processes (MDPs), where the environment setup is inherently Markovian~\cite{mdp}. An MDP is characterized by a set of states $\mathcal{S}$, a set of actions $\mathcal{A}$, a transition function $P[s_{t+1} | s_t, a_t]$ that determines the probability of transitioning from state $s_t$ to state $s_{t+1}$ after taking action $a_t$, and a reward function $r(s_t, a_t)$ which assigns rewards to state-action pairs. The fundamental assumption in an MDP is that the current state encapsulates all necessary information for decision-making, implying that the future state $s_{t+1}$ is conditionally independent of past states given the current state $s_t$ and action $a_t$. Formally, this is expressed as: 
\begin{equation}
P[s_{t+1} | s_t, a_t] = P[s_{t+1} | s_t, a_t, s_{t-1}, a_{t-1}, \ldots, s_0, a_0].    
\end{equation}

Partially Observable Markov Decision Processes (POMDPs) extend the MDP framework to scenarios with partial observability~\cite{pomdp}. In POMDP, at each time step $t$ the agent receives an observation $o_t \in \Omega$, which provides partial information about the actual state $s_t$. The challenge in POMDPs lies in the agent's need to infer the hidden state of the environment from the history of observations and actions. This is a significant complication over the full observability assumption in MDPs, where the current state encapsulates all necessary information for decision-making.

To address the challenges posed by partial observability, strategies often involve utilizing the history of observations and actions or an encoding of this history to approximate the unobservable states, effectively transforming the problem back into a solvable MDP. The agent observes at each time step an encoding of the history $h_t$, from which it needs to infer the hidden state of the environment. The history encoding $h_t$ can be expressed as:
\begin{equation}
h_t = f(s_{t-1}, a_{t-1}, \ldots, s_0, a_0)
\end{equation}
where $f$ represents a function that encodes the history of states and actions into a format that the agent can use for its decision-making process. This encoding can be the stacking of a predefined number of the most recent observations and actions~\cite{atari1, atari2}, or it can be achieved using a Recurrent Neural Network (RNN)~\cite{rdpg1, rdpg2}, which is capable of maintaining and updating internal state representations based on the sequence of observations and actions.

\subsection{Domain Randomization}
\label{subsec:dr}
The challenge of transferring the trained machine learning models, particularly in RL, from simulation to the real world is a major challenge in robotics and control systems~\cite{sim2real-blind}. This challenge, often referred to as the sim-to-real transfer problem, arises because the transition probabilities in a simulated environment, $P_{\textit{sim}}[s_{t + 1} | s_t, a_t]$, do not perfectly match those in the real-world, $P_{\textit{real-world}}[s_{t + 1} | s_t, a_t]$. The discrepancy between these transition probabilities usually leads to a model that performs well in simulation but fails to generalize in real-world conditions.

Domain randomization is a technique designed to address this discrepancy by training the agent in a variety of simulated environments with randomly altered physics parameters, sensor noise, and other environmental conditions. The core idea is to expose the agent to a wide range of possible conditions during the training phase, which helps in improving its ability to generalize from the simulated environment to the real world~\cite{dr-nn-image, dr-sim-to-real}. In standard training, an agent is trained under a fixed dynamics vector $\rho \in \mathcal{P}$, where $\mathcal{P}$ represent the space of all possible dynamics parameters. In domain randomization, for each training episode $i$, a dynamics vector $\rho_i$ is sampled from a predefined distribution over $\mathcal{P}$, i.e. $\rho_i \sim P(\mathcal{P})$. The environment dynamics for that episode are then defined by $\rho_i$, and they remain constant for the duration of the episode. This process can be formalized as follows:

\begin{equation}
    \rho_i \sim P(\mathcal{P}), \quad \forall i \in \{1, 2, \ldots, N\}
\end{equation}

where $N$ is the number of episodes, and $P(\mathcal{P})$ is the probability distribution over the dynamics space $\mathcal{P}$. The goal of domain randomization is to train an agent such that the real-world environment appears as another sample from the distribution $P(\mathcal{P})$. This approach effectively broadens the distribution of the simulation to incorporate a wider range of real-world variations. The agent, therefore, learns a generalized policy $\pi$ that is robust across a variety of dynamics.

\section{Methodology}
\label{sec:methodology}
The optimization of the injection process at COSY currently involves manual adjustments of all sections (1 to 8). This manual method aims to enhance the beam current within COSY but is time-consuming and often results in inconsistent beam properties within the storage ring.

A better alternative strategy is to focus on optimizing the beam's phase space at the point of injection, rather than merely aiming to increase the beam current inside the storage ring. The objective is to align the phase space of the incoming beam with the storage ring's injection acceptance to ensure a consistently high-quality beam. In this research, the agent is assigned to manipulate the 11 magnets (4 quadrupoles and 7 steerers) of section~8  to optimize the transverse space of the beam at the injection point. The settings of sections 1-7 can be chosen independently, as long as 100 \% transmission to the end of section 7 is reached. A possible method to optimize these sections is discussed in \cite{bo}. While acknowledging the importance of the angle of the injected beam, it was excluded from the optimization process due to limitations with the IBL sensors at COSY. The direct feedback on the beam's characteristics is provided by a camera positioned at the end of the IBL. 

\subsection{Explicit Goal}
\begin{figure}[!htb]
   \centering
   \includegraphics*[width=.4\columnwidth]{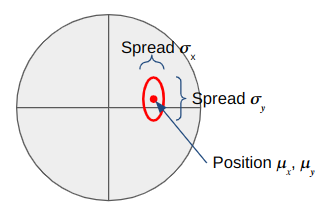}
   \caption{The Optimization goal of the agent is the beam's cross-section at the injection point and is set by operators. It is defined by the center of the beam ($\mu_x$, $\mu_y$) and its spread ($\sigma_x$, $\sigma_y$).} 
   \label{fig:goal}
\end{figure}

The goal $g \in \mathbb{R}^d$ of the optimization is explicitly dictated by the operators. We set the goal as the cross-section (position and width) of the beam at the injection point. Therefore, the goal $g \in \mathbb{R}^4$ is defined as the beam parameters corresponding to the beam center ($\mu_x, \mu_y$) and its spread ($\sigma_x, \sigma_y$) at the injection point as observed through the camera, see Figure~\ref{fig:goal}. This requires the value function $V(s)$ to generalize not only over states, but goals too as a universal value function approximator $V(s, g)$~\cite{goal-policy}. This adjustment demands the introduction of a goal-oriented reward function, $r(s, a, g)$, and modifies the agent's policy correspondingly to $\pi(a | s, g)$. This modification enables the learning of a universal policy adaptable to various target configurations at the injection point by presenting the agent with a randomly sampled goal $g$ at the beginning of each episode. The agent's task is to manipulate the magnets to match the observed beam characteristics with the desired goal, thereby optimizing the transverse position and width of the beam at the point of injection.

\subsection{Randomized simulation dynamics} To train the policy to perform under varying real-world dynamics, the concept of domain randomization is employed in simulation during the training phase. The goal to be achieved here is to train the agent across a multitude of simulated environments, each characterized by distinct physics parameters, sensor noise levels, and environmental factors. This is realized by adjusting the training environment's dynamics through a set of parameters $\rho$ which are varied at the start of each training episode but remain static throughout the same episode. In our experiment, the dynamics vector $\rho$ consists of (1) initial bunch phase space parameters, (2) initial beam angle, offset, and number of clusters, (3) magnets' strength, (4) initial magnets values, and (5) offset to magnets reading value. 

\begin{figure}[!htb]
   \centering
   \includegraphics*[width=.7\columnwidth]{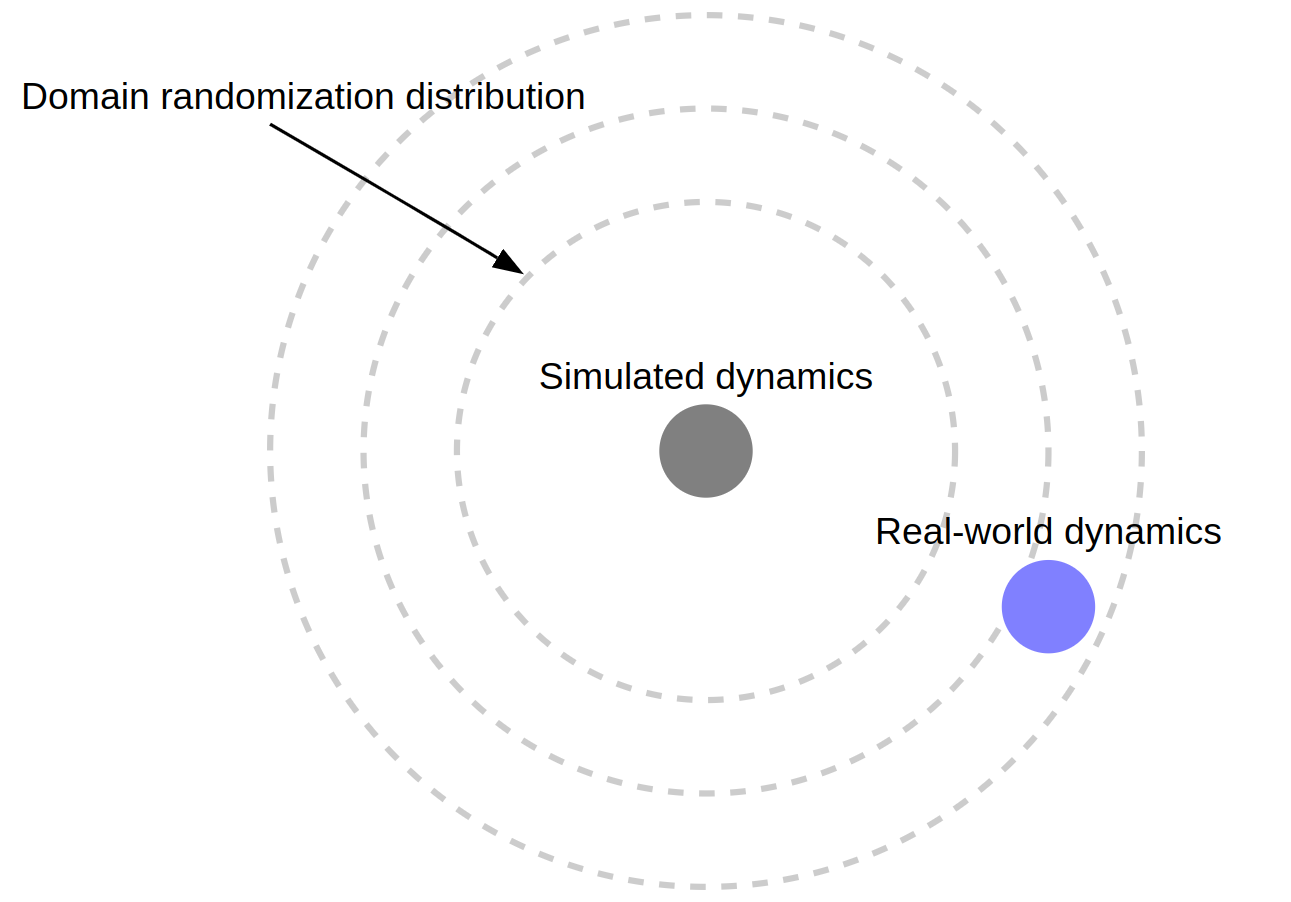}
   \caption{Illustration of the domain randomization concept. Usually there is discrepancy between the simulated dynamics and the real-world dynamics shown here as the grey point and the blue point respectively in the space of all possible dynamics. By expanding the distribution of dynamics from which the simulated dynamics are sampled, the trained agent becomes more robust. If the domain randomization is sufficient, the real-world dynamics appear to the agent as another sample of the simulated environments.~\cite{awal-thesis}}
   \label{fig:domain-rand}
\end{figure}

When the agent is trained under sufficient domain randomization and can perform the optimization in simulation successfully, the agent is expected to perform well in the live run. The reason is, for the agent, the live IBL looks as another sample of the environment dynamics it encountered during the training. If the domain randomization is not sufficient, then the defined distribution over the dynamics space should be increased as shown in Figure~\ref{fig:domain-rand}. Formally, this is correct when the defined distribution over the dynamics space $P(\mathcal{P})$ includes the dynamics of the live IBL $\rho_{\textit{live-IBL}}$:

\begin{equation}
    \rho_{\textit{live-IBL}} \in P(\mathcal{P}) \,.
\end{equation}

In addition, a Gaussian noise $\mathcal{N}(0, \sigma^2)$ is added to the observations which are sampled at each time step of the training. This noise enhances the agent's ability to operate under realistic conditions by simulating the measurement inaccuracies and environmental noise. By exposing the agent policy to a broad range of dynamic conditions, it learns to generalize its performance by learning an internal concept of the environment, improving its applicability to the actual conditions of the IBL at COSY. 

\subsection{Reinforcement Learning Agent} 

The RL agent used in this study is based on the Soft Actor-Critic (SAC) algorithm~\cite{sac}, a model-free actor-critic method \cite{actor-critic, actor-critic2} that is effective in environments requiring continuous action decisions. SAC is distinguished by its incorporation of entropy into the reward structure, promoting exploration by the agent and preventing premature convergence to sub-optimal policies. Based on the earlier discussions, explicit goals $g$, encoded history $h$, and simulation dynamics $\rho$ are incorporated into the policy $\pi_\phi(a|s, h, g)$ and the Q-function $Q_{\theta}(s, h, g, a, \rho)$ which are parameterized using neural networks. Here $\phi$ and $\theta$ denote the weights of the policy network and the Q-function network respectively.

The SAC algorithm optimizes the stochastic policy during off-policy learning, which means it can learn from experiences generated by any policy, not just the current one. The policy network $\pi_\phi$ aims to maximize the expected return as well as the entropy of the policy to encourage exploration. The objective function for the policy network, parameterized by $\phi$, is given by:

\begin{equation}
J(\pi_\phi) = \mathbb{E}_{(s, h, g) \sim \mathcal{D}, a \sim \pi_\phi} \left[ Q_{\theta}(s, h, g, a, \rho) - \alpha \log \pi_\phi(a | s, h, g) \right]
\end{equation}

where $\alpha$ is the temperature parameter that determines the relative importance of the entropy term against the reward, and $\mathcal{D}$ represents the experience replay buffer. The environment dynamics $\rho$ and the goal $g$ are fixed throughout the episode while the others are time step dependent.

The Q-function network $Q_{\theta}(s, h, g, a, \rho)$ is optimized to improve the computed predicted returns, also called the Q-value, by minimizing the difference between the predicted Q-value and the target Q-value, a better estimate than the predicted one. The target Q-value is computed as:

\begin{equation}
    y = \mathbb{E}_{(s, h, g, a, r, s', h') \sim \mathcal{D}} \left[ r + \gamma \left( Q_{\theta}(s', h', g, a', \rho) - \alpha \log \pi_\phi(a'|s', h', g) \right) \right] \, ,
\end{equation}

where $\gamma$ is the discount factor, and unlike the next state $s'$ and next encoded history $h'$, which are sampled from the replay buffer $\mathcal{D}$, the next action $a'$ is computed using the current policy network $\pi_\phi$. 

The Q-network is trained similar to supervised learning using the computed target Q-value. The loss function for optimizing the Q-function network is expressed as:

\begin{equation}
    \mathcal{L}_Q(\theta) = \mathbb{E}_{(s, h, g, a) \sim \mathcal{D}} \left[ \frac{1}{2} \Big( Q_{\theta_i}(s, h, g, a, \rho) - y \Big)^2 \right] \, .
\end{equation}

The SAC algorithm iteratively updates the policy $\pi_\phi$ and the action-value function $Q_\theta$ using samples drawn from the experience replay buffer $\mathcal{D}$. The replay buffer allows the agent to learn from a diverse set of experiences, leading to a robust and generalized policy that can effectively adapt to the variant conditions under the randomized simulation dynamics.

The Q-function network is used only during the training process and is discarded later during the live run. During the training, it implicitly learns the model of the environment and its structure of rewards. This knowledge is then used to compute the gradients with respect to the action that results in a higher reward. Passing the environment dynamics $\rho$ to the action-value function can improve the training by speeding up the convergence of the action-value function and might improve the performance of the agent by computing better gradients of the expected value with respect to the action. The environment dynamics $\rho$ can be, nevertheless, omitted from the action-value function as they can be inferred from the encoded history $h$. This can extend the training time, but the learned policy of the agent is eventually the same.

\subsection{Dense Neural Networks} 
\begin{figure}[!htb]
   \centering
   \includegraphics*[width=.35\columnwidth]{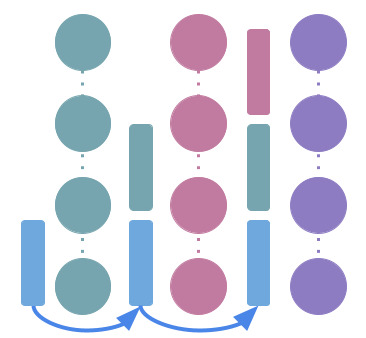}
   \caption{Illustration of the structure of dense neural networks. The colored circles are the neurons and the color matching rectangles are the features generated by the neurons while the blue rectangle represents the input features. Each layer receives the combination of all earlier features in addition to the input features.~\cite{awal-thesis}} 
   \label{fig:dense}
\end{figure}

The choice of the neural network architecture can have a crucial role in the performance of RL agent. Sinha~\emph{et~al.}~\cite{d2rl} showed that better benchmarks can be achieved by incorporating the state into the inputs of each layer of the policy neural network and the state-action into the inputs of each layer of the action-value neural network. We adopted an architecture that closely resembles the original DenseNet architecture~\cite{densenet}, which enhances the flow of information and gradients throughout the network by connecting each layer of the neural network to all earlier layers. This is achieved by computing the inputs $\mathbf{x}$ of each layer $k$ as the concatenation of the non-linear outputs and the inputs of the previous layer as follows:
\begin{equation}
    \mathbf{x}_{k} = [\sigma(\mathbf{z}_{k}); \mathbf{x}_{k-1}] 
\end{equation}
where $\sigma$ is the non-linear operation, $\mathbf{z}$ is the linear transformation of the previous layer, and $\mathbf{x}_0$ are the initial inputs to the neural network. Another way to perceive this architecture, as shown in Figure~\ref{fig:dense}, is that the features passed to each layer are the concatenation of the features generated from all earlier layers in addition to the input features. This architecture facilitates a more effective learning process by ensuring that both low-level features and high-level representations contribute directly to the output of each layer, thus enabling the RL agent to distinguish and utilize complex patterns in the environment more efficiently~\cite{dnet-visual}.

\subsection{Reward Function}
The reward function is a crucial component of the reinforcement learning framework. It guides the agent towards achieving the optimization goal and it can also reinforce or suppress certain behaviors. It evaluates the performance of the agent at each time step $t$ and provides a scalar feedback signal. The reward function used in our approach is composed of several parts, designed to encapsulate different aspects of the beam optimization process.

Firstly, we define a term to quantify the accuracy of the beam's transverse space alignment with the desired parameters at the injection point. This is achieved by computing the difference between the measured beam parameters at the injection point ($\mu'_x, \mu'_y, \sigma'_x, \sigma'_y$) and the target parameters ($\mu_x, \mu_y, \sigma_x, \sigma_y$) using the \emph{softplus} function to ensure a smooth and differentiable measure:

\begin{equation}
    \mathrm{transverse}_t = \mathrm{softplus}\left( \sum_{i \in \{x, y\}} \Big[ |\mu_i - \mu'_i| + |\sigma_i - \sigma'_i| \Big] \right)
\end{equation}

where the softplus function is defined as $\mathrm{softplus}(x) = \log(1+e^x)$. This term aims to minimize the discrepancy between the actual and desired beam parameters, encouraging precise control over the beam's alignment and shape at the point of injection. However, merely using this measure, especially in a complex IBL with high dimensionality, can lead to undesired behavior where the agent loses some or most of the beam but still manages to match the desired transverse space. In our earlier experiments, this undesired behavior was present when only this measure was used. To enforce maintaining the efficiency of beam transmission through the IBL, we introduce a measure that evaluates the fraction of the beam successfully transmitted through the IBL:

\begin{equation}
    \mathrm{transmission}_t = \frac{1}{2n}\sum_{i=1}^n \left[ \hat{d}_i + \mathds{1}_{(\hat{d}_i = 1)} \right]
\end{equation}

where $d_i$ represents the distance traveled by the $i$-th simulated particle, $\hat{x}$ is the scaled variable of $x$ to be within the range $[0,1]$, $n$ is the total number of particles, and $\mathds{1}_{(\textit{condition})}$ is the indicator function, evaluating to \SI{1} when \emph{condition} is met and \SI{0} otherwise. The precision term $\mathrm{precision}_t$ combines the transverse alignment and transmission efficiency, reflecting the overall optimization performance at each time step:

\begin{equation}
    \mathrm{precision}_t = (1 - \widehat{\mathrm{transverse}}_{t}) \cdot \mathrm{transmission}_t
\end{equation}

This measure serves as the primary component of the reward for the agent, promoting the alignment of the beam's transverse space with the desired parameters while ensuring efficient transmission. To discourage unnecessary adjustments and promote stability, a penalty term is introduced. This term penalizes changes in magnet settings that result in a decrease in optimization performance, incorporating a constant $k$ to adjust the penalty's severity:

\begin{equation}
    \mathrm{penalty}_t = k * \mathrm{precision}_t * \mathds{1}_{(m_t \neq m_{t-1})} \mathds{1}_{(\mathrm{precision}_t < r_{t-1})}
\end{equation}

where $m_t$ represents the magnet settings at time $t$. The final reward at time $t$, $r_t$, is then calculated as the precision measure scaled to be negative~\cite{neg-r-effect, neg-r-conceptnn} and adjusted for any penalties incurred:

\begin{equation}
    r_t = \mathrm{precision}_t - \mathrm{penalty}_t - 1
\end{equation}

This reward structure is designed to finely balance the trade-off between the accuracy of beam parameters, efficiency of transmission, and the cost of adjustments, guiding the agent towards achieving the optimal set of actions for beam injection optimization at COSY.

\section{Implementation and Training}
\label{sec:imp-training}
The implementation and training of our RL framework for optimizing the injection process at COSY is based on a simulation environment that mirrors different variants of the real-world dynamics of the IBL. This simulation is the foundation for training and evaluation of the RL agent, enabling it to learn and adapt to the complexities of particle acceleration and beam optimization. At each step of the optimization run, the RL agent is required to determine the adjustments to the magnets based on its current policy. The adjustment is represented by the action $a_t$ at time step $t$, which directly influences the current values of the magnets $m_t$. The new magnet values are calculated using the equation:
\begin{equation}
    m_t = m_{t-1} + a_{t}
\end{equation}

where $m_{t-1}$ represents the magnet values at the previous time step. This formulation of how the new magnet values $m_t$ are updated reflects the practical approach used by operators who use adjustments and incremental changes to manage the dynamic and interdependent nature of the magnets. This approach provides incremental and controlled modifications that ensures maintaining stability and performance without abrupt changes that could lead to beam loss. Additionally, using adjustments allow the RL agent to respond flexibly to the constantly changing conditions and feedback from the system.

It is crucial to maintain the magnet values within operational limits to ensure the safety and integrity of the accelerator's components. Therefore, the magnet values are constrained to operate within the range of $[-0.7, 0.7]$, corresponding to $\pm70\%$ of their maximum current capability. This constraint not only ensures the operational safety of the IBL but also reflects the physical limitations and operational protocols of COSY. Training within these constraints allows the RL agent to develop strategies that are viable for live deployment, ensuring that the optimization strategies are both effective and practical for real-world application.

\subsection{Simulation Environment}
The simulation for training the agent was conducted using MAD-X~\cite{madx}, a comprehensive tool for designing and simulating particle accelerators. A virtual environment was created following the standard Gym interface to simulate the dynamics of the IBL at COSY~\cite{gym, gymnasium}. Particles are simulated and tracked via PTC for more comprehensive analysis~\cite{madx-ptc,madx-ptc2}. This environment simulates the physics of beam propagation through the IBL and models the effects of the quadrupole and steerer magnets on the beam's trajectory and characteristics.

\subsection{Reinforcement Learning Implementation}
The RL model consists of two primary components: the policy network $\pi$ and the action-value network $Q$. Both networks are implemented with dense layers, each with two hidden layers of 512 neurons. The input to the policy network includes the values of the quadrupoles and steerers in the last section of the IBL, computed statistics from the camera output ($\mu_x$, $\mu_y$, $\sigma_x$, $\sigma_y$), the last seven observations of magnet values and their corresponding beam statistics, and the target beam parameters. The $Q$ function receives the same inputs as the policy, with the addition of the environment parameters $\rho$, representing the dynamics of the simulation.

The training process utilized 8 parallel environments to enhance learning stability and efficiency, requiring 8 CPUs for environment simulation and an additional GPU for the agent's processing and neural network training. At each step, a beam simulation with 1000 particles was carried out. Each episode consisted of 32 steps, with the action of the agent having a maximum change of 15\% in the magnet values per step. The discount factor $\gamma$ was set to 0.95, reflecting an emphasis on future rewards. The neural networks are optimized via ADAM optimizer~\cite{adam} with a learning rate of 0.0003. While training was initially set for 10'000 epochs, it was typically terminated earlier upon convergence. For the empirical analysis conducted during the development phase, agents were trained using 4 parallel environments, utilizing 4 CPUs for environment simulation and an additional CPU for agent processing.

\subsection{Live Run Setup}
For the live run, an interface to the IBL was created using the same Gym environment interface, allowing for a seamless transition from simulation to real-world application. The IBL was autonomously controlled using the EPICS control system~\cite{epics1, epics2, cpymad}, enabling real-time adjustments to the magnet settings based on the agent's decisions.

A fluorescent screen and a camera were deployed at the end of the IBL to provide direct feedback on the beam's transverse space. Live image processing was implemented using the area-detector module of the EPICS control system~\cite{epics-areadetector}. The camera, operating at a 100ms frame rate, captures the beam's cross-section at the injection point and provides real-time data on the beam's characteristics. However, due to the release of particles in 20ms bunches, the observed values of $\sigma_x$ and $\sigma_y$ are notably noisy, presenting an additional challenge for the optimization task. 

The agent was evaluated in several different optimization tasks, each requiring adjustments to the beam parameters to meet specific targets. For the live optimization runs, the agent was limited to 16 steps per task.

\section{Experiments and Results}
\label{sec:experiments}
The experiments conducted aimed to validate the effectiveness of the proposed RL approach in optimizing the IBL at COSY. The successful training of the agent across a diverse set of environmental dynamics in simulation, enables it to efficiently optimize the IBL during live runs. For the agent, the real-world IBL is presented as another variant of the various environment dynamics it encountered during the training and can be optimized via the learned robust policy of the trained agent.

\subsection{Live Run Performance}
\begin{figure}[htb]
   \centering
   \begin{tikzpicture}
   \draw(0,0) node{\includegraphics*[width=.5\textwidth]{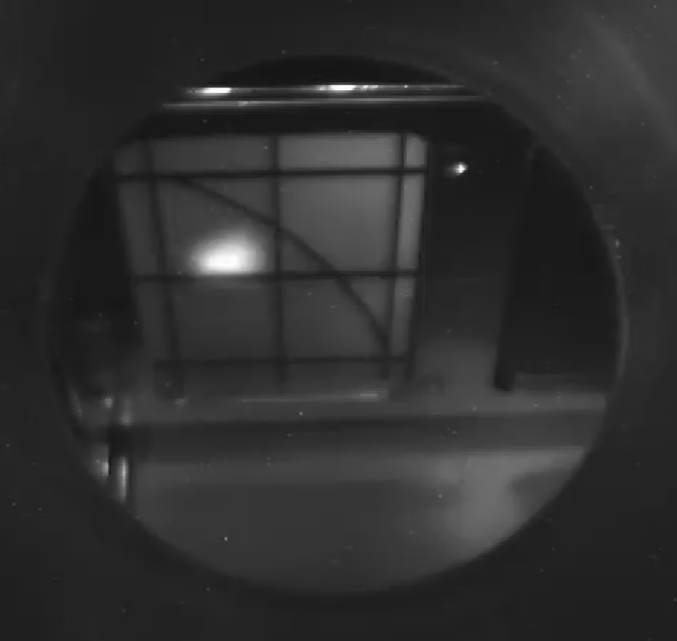}};
   \draw[yellow,|<->|] (-0.65,1.2) --++ (1.25,0) node[midway,above]{1 cm};
   \def\xcenter{-1.3}
   \def\ycenter{.635}
   \draw[red,|-|] (\xcenter-0.1875,\ycenter) -- (\xcenter+0.1875,\ycenter);
   \draw[red,|-|] (\xcenter,\ycenter-0.125) -- (\xcenter,\ycenter+0.125);
   \draw[yellow,->] (2.7,-2.5) --++ (-.5,0) node[midway,above]{$x$};
   \draw[yellow,->] (2.7,-2.5) --++ (0,-.5) node[midway,right]{$y$};
   \end{tikzpicture}
   \caption{A picture of the beam, illuminating through the fluorescent screen, during a real-time optimization by the RL agent. This screenshot was taken during the RL agent optimizing the beam at step~7 of the full 16 steps. The target in this instance for the beam cross-section is shown in red and has the coordinates $\mu_x =$ \SI{5.5}{mm}, $\mu_y =$ \SI{-1.8}{mm}, $\sigma_x =$ \SI{3}{mm}, and $\sigma_y =$ \SI{2}{mm}.}
   \label{fig:live}
\end{figure}

In the live runs, the agent was tasked with optimizing the beam's central position and its spread to meet a set of predefined tasks. These tasks involved positioning the beam at various locations on the fluorescent screen, including the center and the edge, while changing the horizontal and vertical spreads of the beam between \SI{2}{mm} and \SI{5}{mm}. The performance of the agent was quantified using the mean L2 error for the beam's central position ($\mu$) and the mean L1 error for its spread ($\sigma$). The agent successfully optimized the IBL in the live runs achieving a mean L2 error of \SI{0.19}{mm} for the beam's central position and an average L1 error of \SI{0.7}{mm} for its spread while maintaining a 100\% transmission of the incoming beam. Figure~\ref{fig:live} shows a photograph of the beam as it propagates through the fluorescent screen during a live run optimization.

% \begin{figure}[htb]
%    \centering
%    \begin{tikzpicture}
%    \draw(0,0) node{\includegraphics*[width=.5\textwidth]{real_opt.png}};
%    \draw[yellow,|<->|] (-0.55,1.2) --++ (1.25,0) node[midway,above]{1 cm};
%    \draw[red,-] (-.0975,0.36) --++ (.36,0);
%    \draw[red,-] (.085,.1735) --++ (0,.36);
%    \end{tikzpicture}
%    \caption{A picture of the beam, illuminating through the fluorescent screen, during a real-time optimization by the RL agent. The target coordinates for the beam cross-section, shown in red arrows, were $\mu_x =$ \SI{-6.5}{mm}, $\mu_y =$ \SI{2.3}{mm}, $\sigma_x =$ \SI{3}{mm}, and $\sigma_y =$ \SI{3}{mm}. Note that the coordinates are flipped because the image is mirrored.}
%    \label{fig:live}
% \end{figure}

The performance of the RL agent was benchmarked against that of a human operator to evaluate its optimization efficiency. Both the agent and the operator were given a set of optimization tasks to adjust the position of the beam using the steerers only. The operator and the agent achieved a similar optimization accuracy, see Figure~\ref{fig:components}. However, the agent completed the optimization tasks in around \SI{17}{min} while the human operator required around one hour to complete the same tasks. While these tasks were limited to controlling the steerers, the positioning of the magnets within the last section resulted in additional complexity. This further complexity is mainly due to the placement of the last steerers before the last dipole and quadrupole resulting in non-linear dependency of the position on the tuning.

\begin{figure}[ht]
   \centering
   \includegraphics*[width=.75\textwidth]{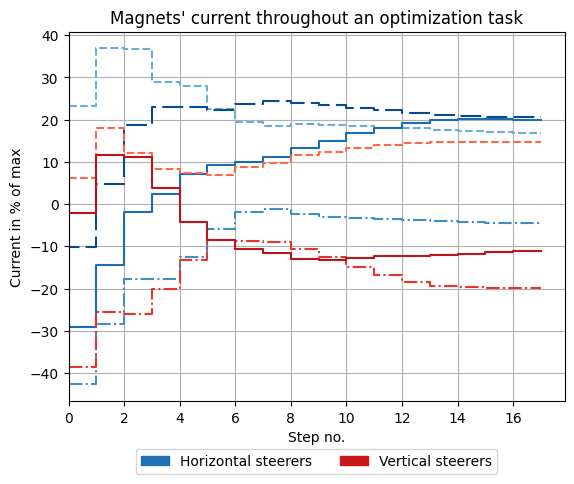}
   \caption{The evolution of the steerers' current throughout a single optimization task induced by the agent actions. This plot illustrates a common optimization pattern where the agent starts with targeted exploration actions and gradually converges to the optimal values as it refines its domain knowledge. Each step is 6 seconds long.}
   \label{fig:steerers}
\end{figure}

Figure~\ref{fig:steerers} provides a visual representation of the optimization process in action, showing the dynamic adjustments made by the agent to the steerers' current during a single optimization task. This visualization illustrates the agent's strategic approach to exploration and optimization. Initially, the agent takes targeted exploration actions and gathers information about the environment from the responses. As the optimization progresses, the agent utilizes this acquired knowledge to make more informed decisions, gradually refining its actions towards the optimal settings. This pattern of behavior illustrates the agent's ability to adapt and optimize in a complex and dynamic environment. 

These results emphasize the potential of RL in enhancing the operational efficiency of particle accelerators. The degree of accuracy achieved by the agent in aligning the beam's parameters with the target values demonstrates the feasibility of employing RL for complex optimization tasks in particle acceleration facilities.

\subsection{Empirical Study on The Architecture Components}
\begin{figure}[ht]
   \centering
   \includegraphics*[width=1.\columnwidth]{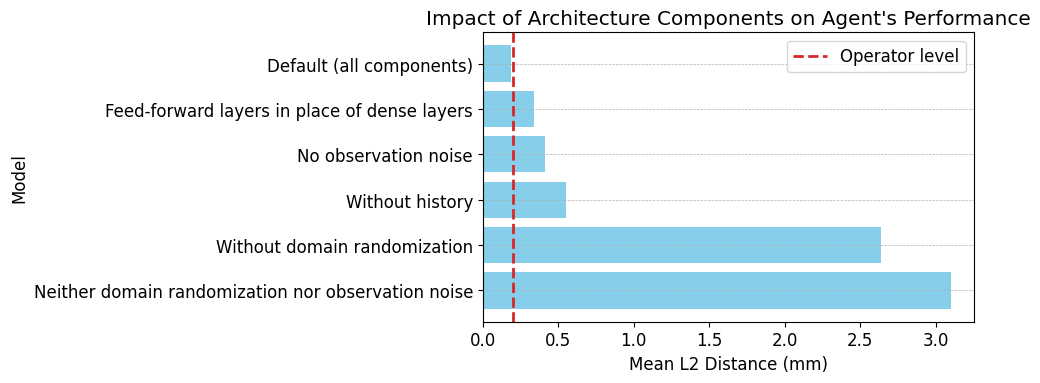}
   \caption{Empirical analysis of the impact of architecture components on the agent's performance. The table presents the mean L2 distance (in mm) between the optimized beam's central position and the target position under various architectural configurations. The "Default" configuration includes all components: dense layers, observation noise, history, and domain randomization. Each subsequent bar shows the performance impact when a specific component is removed or altered, highlighting the importance of each in achieving optimal performance in the beam optimization task. The vertical line marks the optimization level of a human operator.}
   \label{fig:components}
\end{figure}

To further evaluate the impact of the architecture components on the agent's performance, an empirical study was conducted. In this study, the task was limited to controlling the beam location by optimizing the seven steerers in section~8 of the IBL to meet a series of five predefined tasks. The performance of several replicate agents, each lacking a different component of the architecture, was analyzed and compared. The examined components  were the dense layers, observation noise, history, and domain randomization. The results of this study are summarized in Figure~\ref{fig:components}.

The removal of domain randomization had the most significant impact on performance, indicating the critical role of training under varied dynamics for generalizing to real-world conditions. This reinforces the importance of domain randomization in preparing the agent for the complexities and variabilities of the live environment at COSY. The absence of observation noise, while less impactful, still resulted in a noticeable degradation in performance, underlining the importance of training under conditions that mimic the real operational environment, including sensor noise and measurement inaccuracies.

Without the inclusion of history in the agent's observation space, its ability to infer the environment's dynamics and predict the outcome of actions under partial observability was notably hindered. This limitation is evident in the increased error and illustrates the importance of temporal information for making informed decisions in environments where the current state alone does not provide complete information about the system. Utilizing recent advances in machine learning research, by adopting dense layers, allowed the agent to take more precise actions by forming more complex representations of the observations.

This empirical study demonstrates that each component of the architecture contributes to the overall effectiveness of the RL approach in optimizing the IBL at COSY. The integration of dense layers, observation noise, history, and domain randomization into the RL framework is essential for achieving high performance in complex optimization tasks.

\section{Summary and Conclusions}

This research introduces a framework for the application of Reinforcement Learning (RL) to optimize the injection process at accelerator facilities. By employing an RL agent tailored to the specific challenges of optimizing the injection beam line, we developed an agent trained solely in simulation that is capable of optimizing the beam's cross-section dynamically to meet predefined targets during a live operation at the COSY facility.

Live run evaluations demonstrated that the RL agent could effectively optimize the beam's cross-section by aligning the beam's central position and spread with the target values given to the agent by operators. The agent's optimization accuracy was on par with that of a human operator, yet it completed the tasks in a time reduced by a factor of 3 of the time required by the human operator. 

The empirical study of the architecture components validated the importance and significance of each element in the RL framework. Especially,  the results highlighted the critical role of domain randomization and observation noise in preparing an agent trained solely in simulation for real-world control operation. In addition, incorporating historical data into the training process was shown to be essential for dealing with partial observability and enhancing the agent's decision-making capabilities.

In conclusion, this research illustrates the potential of machine learning methods, namely RL, to enhance the efficiency of particle accelerator operations. The successful application of an RL agent for beam injection control at COSY demonstrates the potentials in adopting broader applications of machine learning techniques in particle accelerators and other fields requiring efficient and precise control of complex systems. 

\subsubsection*{Acknowledgement}

The authors thank the COSY crew, particularly V. Kamerdzhiev, for their great assistance in preparation and during the beam shifts, and for the support from JEDI team. In addition, the authors thank Hans-Joachim Stein for the insightful and helpful discussions. Simulations were performed with computing resources granted by RWTH Aachen University under project rwth0905.

\bibliographystyle{unsrt}
%\bibliography{references}
%\printbibliography

\end{document}